\documentclass[fleqn,twoside]{article}
\usepackage{espcrc2}
\usepackage{epsfig}
\usepackage{amssymb}
\usepackage{amsbsy}
\usepackage{amsfonts}
\usepackage{graphics}
\usepackage{latexsym}
\usepackage{epsf}
\usepackage{graphicx}
\usepackage[figuresright]{rotating}

\newcommand{\pslash}{\not{\hspace{-0.08cm}p}}
\newcommand{\Dlr}{\stackrel{\leftrightarrow}{D}}

\title{\vspace{-2.25cm}
       {\normalsize DESY 03--161}    \\[-0.2cm]
       {\normalsize Edinburgh 2003/15} \\[-0.2cm]
       {\normalsize LU-ITP 2003/017} \\[-0.2cm]
       {\normalsize LTH 593} \\[0.50cm]
Quark spectra and light hadron phenomenology from overlap 
fermions with improved gauge field action%
\thanks{Talks presented by M. G\"urtler, R. Horsley, H. Perlt and T. 
Streuer at Lattice 2003.}}

\author{D. Galletly\address{School of Physics, University of Edinburgh, 
Edinburgh EH9 3JZ, UK},
M. G\"urtler\address{John von Neumann-Institut f\"ur Computing 
NIC, Deutsches Elektronen-Synchrotron DESY,\\ 15738 Zeuthen, Germany},
R. Horsley$^{\rm a}$,
B. Jo\'{o}$^{\rm a}$, A.D. Kennedy$^{\rm a}$,
H. Perlt\address{Institut f\"ur Theoretische Physik, Universit\"at 
Regensburg, 93040 Regensburg, Germany},
B.J. Pendleton$^{\rm a}$,
P.E.L. Rakow\address{Theoretical Physics Division, Department of 
Mathematical Sciences, University of Liverpool,\\ Liverpool L69 3BX, UK},
G. Schierholz$^{\rm b,}$\address{Deutsches Elektronen-Synchrotron  
DESY, 22603 Hamburg, Germany},
A. Schiller\address{Institut f\"ur Theoretische Physik, Universit\"at 
Leipzig, 04109 Leipzig, Germany} and
T. Streuer$^{\rm b,}$\address{Institut f\"ur Theoretische Physik, Freie 
Universit\"at Berlin, 14196 Berlin, Germany}\\[0.8em]
{QCDSF -- UKQCD Collaboration}}

\begin{document}

\begin{abstract}
We present first results from a simulation of quenched overlap fermions with
improved gauge field action. Among the quantities we study are the
spectral properties of the overlap operator, the chiral condensate and
topological charge, quark and hadron masses, and selected nucleon matrix 
elements. To make contact with continuum physics, we compute the 
renormalization constants of quark bilinear operators in perturbation theory
and beyond.
\end{abstract}

\maketitle

\section{INTRODUCTION}

Lattice calculations at small quark masses, i.e. in the chiral regime,  
require actions with good chiral properties. Overlap fermions~\cite{Neuberger}
have an exact chiral symmetry on the lattice~\cite{Luescher} and thus are 
predestined for this task. A further advantage of overlap fermions is that
they are automatically $O(a)$ improved~\cite{QCDSF1}.


The massive overlap operator is defined by
\begin{equation}
D=\big(1-\frac{am_q}{2\rho}\big)\, D_N + m_q ,
\end{equation}
\begin{equation}
D_N=\frac{\rho}{a}\big(1+\frac{X}{\sqrt{X^\dagger X}}\big),
X=D_W-\frac{\rho}{a},
\label{DN}
\end{equation}
where $D_W$ is the Wilson-Dirac operator. We assume $r=1$ throughout this 
paper. The operator $D_N$ has $n_- + n_+$
exact zero modes, $D_N \psi_n = 0$, $n_-$ ($n_+$) being the number of modes 
with negative (positive) chirality, $\gamma_5 \psi_n = - \psi_n$ ($\gamma_5 
\psi_n = + \psi_n$). The index of $D_N$ is thus given by $\nu = n_- - n_+$.
The `continuous' modes $\lambda$, $D_N \psi_\lambda = \lambda \psi_\lambda$, 
having $(\psi_\lambda^\dagger,\gamma_5 \psi_\lambda^{}) = 0$, come in complex 
conjugate pairs $\lambda, \lambda^*$.

To compute the `sign function'
\begin{equation}
{\rm sgn}(X) = \frac{X}{\sqrt{X^\dagger X}} \equiv 
\gamma_5\, {\rm sgn}(H),\, H = \gamma_5 X, 
\end{equation}
we use Zolotarev's optimal rational 
approximation~\cite{Lippert}. To improve the accuracy of the rational 
approximation, and to reduce the number of iterations in the inner inversion,
we project out the $\sim 16$ lowest eigenvalues of $H$. The approximation of 
the `sign function' is done to better than $5 \cdot 10^{-7}$ in the interval 
$[0.1,2.4]$. We use a multi-mass conjugate gradient solver in both the inner 
and outer 
inversions. For any given quark mass this allows to compute propagators for 
a whole set of higher quark masses at very little extra cost in CPU time.
For the inner and outer inversions a stopping criterion of $10^{-6}$ and
$5 \cdot 10^{-6}$, respectively, is employed.

It is important to use a good gauge field action, because the inversion time 
of the fermion matrix is greatly reduced for improved gauge field actions. We 
use the L\"uscher-Weisz action~\cite{LW}
\begin{eqnarray}
S[U]&\!\!\!=\!\!\!&\frac{6}{g^2}\Big[ c_0\!\!\!\! 
\sum_{\rm plaquette}\frac{1}{3}\,
\mbox{Re}\, 
\mbox{Tr}\, (1-U_{\rm plaquette}) \nonumber \\
&\!\!\!+\!\!\!& c_1\!\!\!\! \sum_{\rm rectangle}\frac{1}{3}\, \mbox{Re}\, 
\mbox{Tr}\, (1-U_{\rm rectangle}) \label{ImpAct} \\
&\!\!\!+\!\!\!& c_2\!\!\!\!\!\!\!\! \sum_{\rm parallelogram}\frac{1}{3}\, 
\mbox{Re}\, \mbox{Tr}\, (1-U_{\rm parallelogram}) \Big] \nonumber
\end{eqnarray}
with coefficients $c_1$, $c_2$ ($c_0 + 8 c_1 + 8 c_2 = 1$) taken from tadpole 
improved perturbation theory~\cite{Gattringer}.
In Fig.~1 we compare the condition number of the
Wilson and tadpole improved L\"uscher-Weisz gauge field action. We see that
the condition number is a factor of $\gtrsim 3$ larger for the Wilson action. 
This is largely due to the fact that the L\"uscher-Weisz action suppresses
dislocations~\cite{GKLSW} and thus greatly reduces the number of (unphysical) 
zero modes. Topological studies using the Wilson gauge field action should be
taken with caution. 

\begin{figure}[t]
\begin{center}
\epsfig{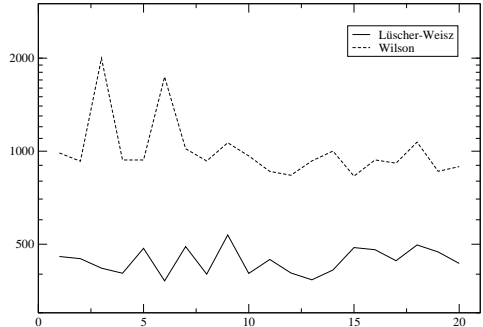}
\vspace*{-0.5cm}
\caption{The condition number for the Wilson and tadpole improved 
L\"uscher-Weisz gauge field action on the $16^3\,32$ lattice at lattice 
spacing $a \approx 0.1$ fm for $\rho=1.4$, as a function of configuration
number.}
\end{center}
\vspace*{-1cm}
\end{figure}

The calculations are mainly done on the $16^3\, 32$ lattice at 
$\beta = (6/g^2)\,c_0 = 8.45$, where~\cite{Gattringer} $c_1 = -0.15486$ and 
$c_2=-0.013407$. The corresponding lattice spacing is $a=0.095$ fm if we use 
$r_0 = 0.5$ fm to set the scale. We have taken $\rho = 1.4$, which we have 
found to be the optimal choice. In addition, part of the calculations have
been done on the $12^3\,24$ lattice at $\beta=8.1$ to test for scaling. The 
lattice spacing here is $a=0.125$ fm~\cite{Gattringer}. Both lattices have 
the same physical volume $V$. 

\section{SPECTRAL PROPERTIES}

The eigenvalues of $D_N$ lie on a circle around ($\rho$,0) with radius 
$\rho$ in the complex plane. The improved operator 
$D_N^{\rm imp} = (1-aD_N/2\rho)^{-1} D_N$ \cite{QCDSF1} projects the 
eigenvalues of $D_N$ stereographically onto the imaginary axis.
The `continuous' eigenvalues of $D_N^{\rm imp}$ come in pairs $\pm {\rm i} 
\lambda$, while the zero modes are untouched.  

We have computed the lowest $\approx$ 140 eigenvalues of $D_N$ on both
our lattices for $O(250)$ gauge field configurations each. We employed the 
Arnoldi algorithm as provided by the ARPACK package.

If one defines the topological charge density by $q(x)=(1/2){\rm Tr}\,
(\gamma_5 D_N(x,x))$~\cite{Niedermayer}, the total charge $Q$ is given by the 
index of $D_N$:
\begin{equation}
Q=\sum_x q(x) = n_- - n_+ .
\label{charge}
\end{equation}
Using this definition we have computed the topological susceptibility 
$\chi_{\rm{top}} = \langle Q^2\rangle /V$ and find
\begin{equation}
\hspace*{1.75cm}
\begin{tabular}{c|c}
$\beta$ & $\chi_{\rm top}$ \\ \hline
8.10 & $(195(5)\,{\rm MeV})^4$ \\
8.45 & $(187(5)\,{\rm MeV})^4$
\end{tabular}
\end{equation}
 
\begin{figure}[t]
\begin{center}
\epsfig{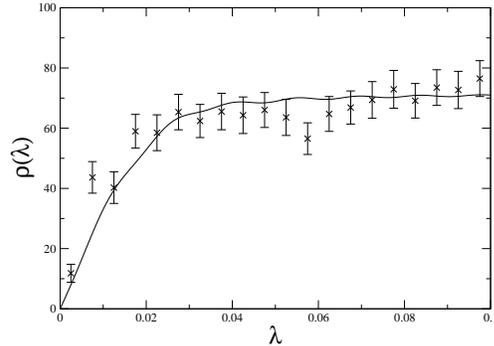}
\vspace*{-0.5cm}
\caption{The spectral density $\rho(\lambda)$ at $\beta=8.45$, together with a
fit of random matrix theory to the data. Both $\rho$ and $\lambda$ are given
in lattice units.}
\end{center}
\vspace*{-1cm}
\label{plot_rho}
\end{figure}

The spectral density of the `continuous' modes is formally given by
\begin{equation}
\rho(\lambda)=\frac{1}{V} \big\langle\sum_{\bar{\lambda}} \delta(\lambda
-\bar{\lambda})\big\rangle ,
\end{equation}
where the sum extends over the (nonzero) eigenvalues $\pm {\rm i}
\bar{\lambda}$ of $D_N^{\rm imp}$. In our case $0 < |\bar{\lambda}| 
\lesssim 800$ MeV. In practice one groups the eigenvalues into bins, whose 
size will depend on the statistics.
In the infinite volume $\Sigma \equiv -\langle\bar{\psi}\psi\rangle = - 
\pi\rho(0)$. In the finite volume and for small eigenvalues the spectral 
density can be computed from the chiral low-energy effective theory. 
For $\lambda < E_T$,  $E_T$ being the Thouless energy $E_T \approx 
f_{\pi}^2/\Sigma \sqrt{V}$, the low-energy effective partition function is 
dominated by the zero momentum modes, and the zero-momentum approximation of 
the chiral low-energy effective theory is equivalent to chiral random matrix 
theory. In random matrix theory the microscopic spectral density in the
sector of fixed topological charge $Q$, 
\begin{equation}
\rho_S^{(Q)}(\Sigma V \lambda) 
\equiv \lim_{V \rightarrow \infty} \frac{1}{\Sigma}\,\rho(\lambda)\,\big|_Q
\end{equation}
with $\Sigma V \lambda$ kept finite, is given by~\cite{RMT}
\begin{equation}
\rho_S^{(Q)}(x)=\frac{x}{2}\big(J_{|Q|}^2(x)-J_{|Q|+1}(x)J_{|Q|-1}(x)\big) ,
\end{equation}
where $J_n(x)$ are Bessel functions.
Thus, for $\lambda < E_T$ the spectral density is given by
\begin{equation}
\rho(\lambda) = \Sigma \sum_Q w(Q)\,\rho_S^{(Q)}(\Sigma V \lambda) ,
\label{rho}
\end{equation}
where $w(Q)$ is the weight of the sector of topological charge Q with
$\sum_Q w(Q) = 1$. Taking $w(Q)$ from our `measured' charge distributions,
we then may obtain $\Sigma$ by fitting (\ref{rho}) to our data.
In Fig.~\ref{plot_rho} we show the data together with the fit for $\beta=8.45$.
We observe that random matrix theory describes the data well up to 
$\lambda \approx 150\, {\rm MeV}$. For the (unrenormalized) chiral condensate
we find $\Sigma=(242(8)\,{\rm MeV})^3$ at $\beta=8.1$ and 
$\Sigma=(249(9)\,{\rm MeV})^3$ at $\beta=8.45$. We will give renormalized
values after we have computed the renormalization constants.

Random matrix theory predicts furthermore the distribution of the smallest 
nonzero eigenvalue in the sector of topological charge $Q$, which we call  
$\rho_{SE}^{(Q)}(\Sigma V \lambda)$. The first few expressions are~\cite{RMT}:
\begin{eqnarray}
\rho_{SE}^{(0)}(x)&=&\frac{x}{2}\,{\rm e}^{-x^2/2} , \nonumber\\
\rho_{SE}^{(1)}(x)&=&\frac{x}{2}\,{\rm e}^{-x^2/2}I_2(x) ,\\
\rho_{SE}^{(2)}(x)&=&\frac{x}{2}\,{\rm e}^{-x^2/2}\big(I_2^2(x)
-I_1(x)I_3(x)\big) .\nonumber
\end{eqnarray}
After having determined $\Sigma$ from the
spectral density of all eigenvalues, these formulae contain no free parameters 
anymore, and thus can be compared directly to our `measured' distributions.
In Fig.~3 we show our data together with the predictions of
random matrix theory. We find good agreement.

\begin{figure*}[t]
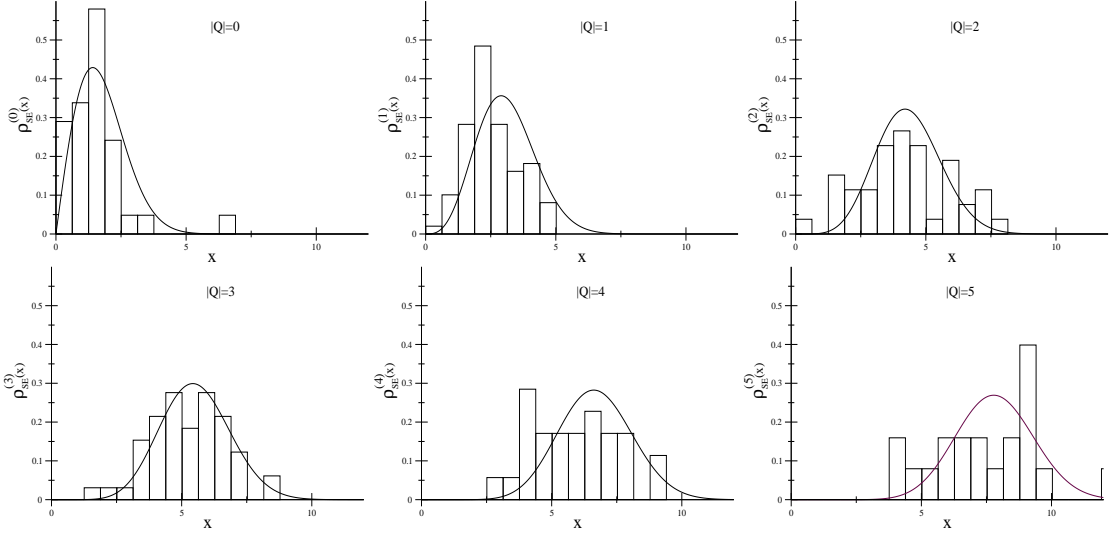

\vspace*{-0.15cm}
\begin{center}
\epsfig{file=e1_0_new.eps,width=4.8cm, height=3.5cm,clip=}
\epsfig{file=e1_1_new.eps,width=4.8cm, height=3.5cm,clip=}
\epsfig{file=e1_2_new.eps,width=4.8cm, height=3.5cm,clip=}
\epsfig{file=e1_3_new.eps,width=4.8cm, height=3.5cm,clip=}
\epsfig{file=e1_4_new.eps,width=4.8cm, height=3.5cm,clip=}
\epsfig{file=e1_5_new.eps,width=4.8cm, height=3.5cm,clip=}
\vspace*{-0.5cm}
\caption{Distributions of the smallest nonzero eigenvalue 
$\rho_{SE}^{(Q)}(x)$ as a function of $x=\Sigma V \lambda$ for 
$|Q|=0,\cdots,5$ on the $16^3\,32$ lattice at $\beta=8.45$, 
together with the predictions of 
random matrix theory, in lattice units.}
\end{center}
\vspace*{-0.35cm}
\label{plot_sm}
\end{figure*}


\section{RENORMALIZATION}
\label{renormalisation}

To obtain physical results from lattice calculations of hadron matrix 
elements the underlying operators have to be renormalized. Let us denote the
lattice regularized operators by $\mathcal{O}(a)$. We then define renormalized
operators $\mathcal{O}^R(\mu)$ by introducing the renormalization constant 
$Z_\mathcal{O}(a\mu)$:
\begin{equation}
\mathcal{O}^R(\mu) = Z_\mathcal{O}(a\mu) \mathcal{O}(a) .
\end{equation}
The renormalization constant $Z_\mathcal{O}(a\mu)$ is found by imposing the 
renormalization condition
\begin{eqnarray}
\Lambda_{\raisebox{-0.05cm}{$\scriptstyle \mathcal{O}$}} \big|_{p^2=\mu^2}
&=& Z_\psi(a\mu) Z_{\raisebox{-0.05cm}{$\scriptstyle \mathcal{O}$}}(a\mu)^{-1}
\Lambda_\mathcal{O}^{{\rm tree}} \nonumber \\
&+& {\rm other\; Dirac\; structures} ,
\label{condition}
\end{eqnarray}
where $Z_\psi$ is the wave function renormalization constant to be defined 
below and $\Lambda_\mathcal{O}$ is the forward (quark line) amputated Green
function computed between off-shell quark states with 4-momentum $p$. We 
consider quark bilinear
operators only. The renormalized operator $\mathcal{O}^R(\mu)$ is independent of
the regularization scheme, but will depend on the external states and on the 
gauge. The operator matrix elements can be converted to more popular schemes 
like $\overline{MS}$ by means of continuum perturbation theory.

In the following we present first results of a perturbative calculation of
renormalization constants for overlap fermions and improved gauge field
action. So far results for overlap fermions are only known for the Wilson 
gauge field action~\cite{Alexandrou,Capitani}. The numbers given below hold
for $c_1=-0.15486$ and $c_2=-0.013407$ (i.e. the improvement coefficients at 
$\beta = 8.45$), while we keep the (bare) coupling constant $g$ arbitrary, 
which allows us to exchange it for a better expansion parameter.

The lattice Feynman rules for overlap fermions have been derived 
in~\cite{Kikukawa,Ishibashi}. The gluon propagator for improved
gauge field action is known in four dimensions 
only~\cite{Weisz,Iwasaki}. A suitable form for dimensional regularization is 
\begin{equation}
D_{\mu\nu}^{\rm imp} =  D_{\mu\nu}^{\rm Wilson} + \Delta D_{\mu\nu} ,
\label{fullprop}
\end{equation}
where 
\begin{equation}
D_{\mu\nu}^{\rm Wilson} = \frac{1}{\hat{k}^2}\left( \delta_{\mu\nu} - \xi\, \frac{\hat{k}_\mu \hat{k}_\nu}{\hat{k}^2}\right) ,
\end{equation}
is the standard Wilson propagator with $\hat{k}_\mu = (2/a)
\sin(ak_\mu/2)$, and $\xi$ specifies the gauge. The Landau gauge corresponds 
to $\xi = 1$, the Feynman gauge to $\xi = 0$. The six-link interaction terms
are contained in $\Delta D_{\mu\nu} = D_{\mu\nu}^{\rm Imp}
-D_{\mu\nu}^{\rm Wilson}$. While $D_{\mu\nu}^{\rm Wilson}$ leads to infrared 
divergent expressions, which have to be regularized, expressions involving 
$\Delta D_{\mu\nu}$ are infrared finite and thus can be computed in four 
dimensions. We may write
\begin{eqnarray}
\Delta D_{\mu\nu } &\!\!\!=\!\!\!&  \delta_{\mu\nu} \sum_{n=0}^4 
D_n(\hat{k},c_1,c_2)\,\hat{k}_\mu^{2n} \nonumber  \\[-0.3em]
& & \label{DeltaD} \\[-0.3em]
&\!\!\!+\!\!\!&  \sum_{m,n=0}^{m+n=4} 
D_{m,n}(\hat{k},c_1,c_2)\,\hat{k}_\mu^{2m+1}\,\hat{k}_\nu^{2n+1} .
\nonumber
\end{eqnarray}
The coefficient functions $D_n$ and $D_{m,n}=D_{n,m}$ are rational functions 
involving $\hat{k}$ and $c_{1,2}$. Explicit expressions are given 
in~\cite{Horsley}. Both functions vanish in the limit $c_{1,2} \rightarrow 0$.

\subsection*{\it Self energy}
\vspace*{0.25cm}

The inverse of the massless quark propagator can be written in the form
\begin{equation}
  S^{-1} = {\rm i}\pslash - \Sigma^{\rm lat}
  \label{qprop}
\end{equation}
with 
\begin{equation}
  \Sigma^{\rm lat} = \frac{g^2\,C_F}{16 \pi^2}\, {\rm i}\pslash \, 
\Sigma_1(a^2p^2) .
\end{equation}
For the quark self energy $\Sigma_1$ we find
\begin{eqnarray}
  \Sigma_1(a^2p^2) &=&  (1-\xi) \, \ln(a^2 p^2) \nonumber \\
 &+& 4.79201 \, \xi + b_\Sigma 
\label{Sigma}
\end{eqnarray}
with
\begin{equation}
b_\Sigma = -16.179 .
\end{equation}
From (\ref{Sigma}) we obtain the quark wave function renormalization constant
\begin{equation}
Z_{\psi}(a\mu) = 1 - \frac{g^2\,C_F}{16 \pi^2}\, \Sigma_1(a^2\mu^2) .
\end{equation}

\subsection*{\it Local operators}
\vspace*{0.25cm}

Let us consider local operators
\begin{equation}
\mathcal{O}_X = \bar{\psi}\, \Gamma^X \psi \equiv X
\end{equation}
now with 
\begin{displaymath}
\hspace*{2.5cm}
\begin{tabular}{c|c}
$X$ & $\Gamma^X$ \\ \hline
$S$ & 1 \\
$P$ & $\gamma_5$ \\
$V$ & $\gamma_\mu$ \\
$A$ & $\gamma_\mu \gamma_5$ \\
$T$ & $\sigma_{\mu\nu} \gamma_5$ \\
\end{tabular}
\end{displaymath}
To find the renormalization constants we have to compute the amputated Green 
functions $\Lambda_{\mathcal{O}_X} \equiv \Lambda^X$. We obtain
\begin{eqnarray}
\Lambda^{S,P} &\!\!\!\!=\!\!\!\!& \{1,\gamma_5\} + \frac{g^2 C_F}{16 \pi^2}
   [- (4-\xi) \, \ln(a^2 p^2) \nonumber \\
&\!\!\!\!-\!\!\!\!& 5.79201 \xi + b_{S,P} ] \, \{1,\gamma_5\} , 
\end{eqnarray}
\begin{eqnarray}
\Lambda_\mu^{V,A} &\!\!\!\!=\!\!\!\!& \{\gamma_\mu,\gamma_\mu \gamma_5\} 
+ \frac{g^2 C_F}{16 \pi^2} \Big\{[ - (1-\xi) \, \ln(a^2 p^2) \nonumber \\
&\!\!\!\!-\!\!\!\!& 4.79201  \xi +   b_{V,A}  ]\gamma_\mu \\
 &\!\!\!\!-\!\!\!\!& 2(1-\xi)\,   \frac{p_\mu \pslash}{p^2}
 \Big\}  \, \{1,\gamma_5\} , \nonumber
\end{eqnarray}
\begin{eqnarray}
\Lambda_{\mu\nu}^T &\!\!\!\!=\!\!\!\!& \sigma_{\mu\nu} \gamma_5 +
\frac{g^2\,C_F}{16 \pi^2} [ \xi\, \ln(a^2 p^2) \nonumber \\
&\!\!\!\!-\!\!\!\!& 3.79201  \xi +  b_{T} ] \sigma_{\mu\nu} \gamma_5 ,
\end{eqnarray}
where
\begin{eqnarray}
b_{S,P} &=& 10.512 , \nonumber \\
b_{V,A} &=& 6.228 , \\
b_T &=& 3.900 . \nonumber
\end{eqnarray}
Using (\ref{condition}) we then arrive at the renormalization constants
\begin{eqnarray}
  Z_{S,P} &\!\! =\!\! &  1 - \frac{g^2 C_F}{16 \pi^2} [-6
  \ln(a\mu)- \xi + b_{S,P} + b_\Sigma  ] , \nonumber \\
  Z_{V,A} & \!\!=\!\! &  1 - \frac{g^2 C_F}{16 \pi^2} \left[b_{V,A} +
  b_\Sigma  \right] ,  \\
  Z_T & \!\!=\!\! &  1 - \frac{g^2 C_F}{16 \pi^2}
  [2\ln(a\mu)+ \xi + b_T +b_\Sigma  ] . \nonumber
\end{eqnarray}
In the $\overline{MS}$ scheme this gives
\begin{eqnarray}
  Z_{S,P}^{\overline{MS}} &\!\! =\!\! &  1 - \frac{g^2 C_F}{16 \pi^2} \left[-6
  \ln(a\mu)- 5 + b_{S,P} + b_\Sigma  \right] ,  \nonumber \\
  Z_{V,A}^{\overline{MS}} & \!\!=\!\! &  1 - \frac{g^2 C_F}{16 \pi^2} \left[b_{V,A} +
  b_\Sigma  \right] ,  \\
  Z_T^{\overline{MS}} & \!\!=\!\! &  1 - \frac{g^2 C_F}{16 \pi^2}
  \left[2\ln(a\mu)+ 1 + b_T +b_\Sigma  \right] . \nonumber
\end{eqnarray}

\subsection*{\it One-link operator}
\vspace*{0.25cm}

The one-link operator that will be of interest to us here is
\begin{equation}
\mathcal{O}_{\mu\nu} = \frac{\rm i}{2} \bar{\psi} \gamma_\mu 
\Dlr_\nu \psi \,\, - \,\, {\rm Traces} .
\end{equation}
Two different irreducible representations of $\mathcal{O}_{\mu\nu}$ under the 
hypercubic group have been 
considered in the literature~\cite{QCDSF3}:
\begin{eqnarray}
\mathcal{O}_a &=& \mathcal{O}_{\{14\}} , \nonumber \\
\mathcal{O}_b &=& \mathcal{O}_{44} -\frac{1}{3}\sum_{i=1}^3{\cal O}_{ii} ,
\end{eqnarray}
where $\{\cdots\}$ denotes symmetrization of the indices. Because of space 
limitations we will not give the Green functions here but only state the final
results. For the renormalization constants of the operators $\mathcal{O}_{a,b}$
we obtain in the $\overline{MS}$ scheme
\begin{equation}
Z_{a,b}^{\overline{MS}} = 1 - \frac{g^2 C_F}{16 \pi^2}\Big[\frac{16}{3}
\ln (a\mu) + b_{a,b} + \frac{40}{9} + b_\Sigma\Big] 
\end{equation}
with
\begin{eqnarray}
b_a & =& -6.516 , \nonumber \\
b_b & =&  -5.617 .
\end{eqnarray}

\subsection*{\it Improvement}
\vspace*{0.25cm}

Lattice perturbation theory is known to converge badly due to the appearance 
of (gluon) tadpole diagrams, which are lattice 
artifacts and which make the bare coupling $g$ into a poor expansion parameter.
It was proposed~\cite{Lepage} that the perturbative series should be 
rearranged in order to get rid of the tadpole contributions. We have made use 
of this observation already in tuning the coefficients of the 
L\"uscher-Weisz action (\ref{ImpAct}). For the renormalization constants this 
rearrangement is done in~\cite{Horsley}, following a method similar to that
in~\cite{QCDSF2}. Writing
\begin{equation}
Z_{\cal O} =  1 - \frac{C_F\, g^2}{16 \pi^2} \tilde{B}_{\cal O} ,
\end{equation}
we arrive at the tadpole improved result
\begin{eqnarray}
Z_{\cal O}^{TI} &=&  \frac{\rho\, u_0^{1-n_D}}{\rho -4+4 u_0} \Big\{ 1 
- \frac{C_F\, g^2}{16 \pi^2\,u_0^4} 
\Big[\tilde{B}_{\cal O} \nonumber \\
&-& \Big(1 - \frac{4}{\rho}-n_D\Big) k_u\Big]\Big\} ,
\end{eqnarray}
where 
\begin{equation}
u_0 = \big\langle \frac{1}{3} {\rm Tr}\, U_{\rm plaquette}
\big\rangle^{\frac{1}{4}} ,
\end{equation}
$n_D$ is the number of covariant derivatives and~\cite{Alford} 
$k_u=0.7325\,\pi^2$. The difference to the Wilson gauge field action is 
that $(1-n_D)\pi^2$ now has to be replaced by $(1-4/\rho-n_D)k_u$, and there
is an additional prefactor of $\rho/(\rho-4-4u_0)$. At our value of
$\beta$ (i.e. $g^2=1.6658$) we find $u_0^4 = 0.65176$.

Alternatively, for the local operators (with $n_D=0$) one may improve the 
perturbative result by writing~\cite{Kronfeld}
\begin{equation}
Z_{\cal O}^{VI} =  Z_V^{\rm nonpert} \Big[1 -
\frac{C_F\, g^2}{16 \pi^2} (\tilde{B}_{\mathcal{O}}-\tilde{B}_V)\Big] ,  
\end{equation}
where $Z_V^{\rm nonpert}$ is the nonperturbatively determined renormalization
constant of the local vector current and $Z_V^{\rm pert}$ the perturbatively
computed one. A similar procedure can be envisaged for the one- and higher-link
operators. In that case $Z_V^{\rm nonpert}$ will have to be replaced by the
appropriate nonperturbatively determined renormalization constant of the
one- and higher-link operator, respectively.

In Table~1 we compare the results of the various
improvement schemes with the perturbative result. For $Z_V^{\rm nonpert}$
we have taken the nonperturbative result $Z_A = 1.416$ derived
below (eq.~(\ref{za})).
\begin{table}[!htb]
\vspace*{-0.25cm}
  \begin{center}
  \begin{tabular}{|c|c|c|l|}
  \hline
  ${\cal O}$ &  $Z_{\cal O}$ &  $Z_{\cal O}^{TI}$ &  
\multicolumn{1}{c|}{$Z_{\cal O}^{VI}$}\\
  \hline
  $S,P$            &   1.150  & 1.190 & 1.430   \\ [0.7ex]
  $V,A$            &   1.140  & 1.171 & 1.416    \\ [0.7ex]
  $T$              &   1.159  & 1.207 & 1.443    \\ [0.7ex]
  ${\cal O}_a$   &   1.257  & 1.335 & \multicolumn{1}{c|}{-}    \\ [0.7ex]
  ${\cal O}_b$   &   1.244  & 1.308 & \multicolumn{1}{c|}{-}    \\ [0.7ex]
  \hline
  \end{tabular}
  \end{center}
  \caption{Comparison of renormalization constants at the scale $a\mu = 1$ for
  $\beta=8.45$ and $Z_V^{\rm nonpert} = 1.416$ in the $\overline{MS}$ scheme.}
  \label{tabZ}
\vspace*{-0.25cm}
\end{table}

\section{CHIRAL CONDENSATE}

Knowing $Z_S$, we can now compute the renormalized chiral condensate
\begin{equation}
\langle \bar{\psi}\psi\rangle^R(\mu) = - Z_S(a\mu)\, \Sigma .
\end{equation}
It is traditional to quote numbers for $\mu = 2$ GeV. In the $\overline{MS}$ 
scheme we obtain $Z_S^{TI}(2\,{\rm GeV}) = 1.253$ at $\beta=8.1$ and
$Z_S^{TI}(2\,{\rm GeV}) = 1.184$ at $\beta=8.45$. This gives
\begin{equation}
\hspace*{1.5cm}
\begin{tabular}{c|c}
$\beta$ & $\langle \bar{\psi}\psi\rangle^{\overline{MS}}(2\,{\rm GeV})$ \\ 
\hline
8.10 & (261(9)\,{\rm MeV})$^3$ \\
8.45 & (263(9)\,{\rm MeV})$^3$
\end{tabular}
\end{equation}
We find good agreement with scaling. 

At $\beta=8.45$ we have $Z_S^{VI}(2\,{\rm GeV}) = 1.426$. Using this value for
the renormalization constant we obtain
$\langle \bar{\psi}\psi\rangle^{\overline{MS}}(2\,{\rm GeV}) =
(280(10)\,{\rm MeV})^3$.

\section{QUARK AND HADRON MASSES}

The results presented in this and the next Section refer to the $16^3\, 32$ 
lattice at $\beta=8.45$. The calculations are done for four different quark 
masses, $am_q = 0.028$, 0.056, 0.098 and 0.140 (corresponding to $a\mu 
\equiv am_q/2\rho = 0.01$, 0.02, 0.035 and 0.05.) To increase the overlap of 
meson and baryon operators with the ground state wave function we use smeared 
sources~\cite{QCDSF4} with $\kappa_s = 0.21$ and $N_s = 50$. The calculations 
of two-point functions are based on $O(100)$ configurations. The code 
has partly been written in SZIN~\cite{szin}, which has the advantage of 
being flexible and machine independent.

\begin{figure}[tbp]
  \centering
  \epsfig{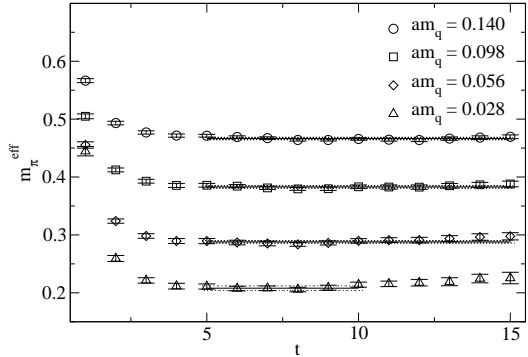}
\vspace*{-0.5cm}
  \caption{Effective pion masses for our four different quark masses from 
 $\langle A_4(t)A_4(0)\rangle$.}
\vspace*{-0.25cm}
  \label{fig:mpi_a4}
\end{figure}

\subsection*{\it Pion mass}
\vspace*{0.25cm}

We compute the pion mass from various correlation functions: $\langle P(t) 
P(0) \rangle$, $\langle A_4(t) P(0) \rangle$ and $\langle A_4(t) A_4(0) 
\rangle$, where $P$ is the pseudoscalar density and $A_\mu$ the axial vector 
current. In Fig.~4 we show the effective mass plot of $\langle A_4(t) A_4(0) 
\rangle$ for our four quark masses. This correlation function is least 
affected by finite size corrections induced by zero mode 
contributions~\cite{Blum,Liu}. 
The resulting pion masses are given in Table~2. Our lightest pion mass is
$\approx 400$ MeV, which is compatible with quenched Wilson fermion 
calculations. 

\begin{figure}[tbp]
  \centering
  \epsfig{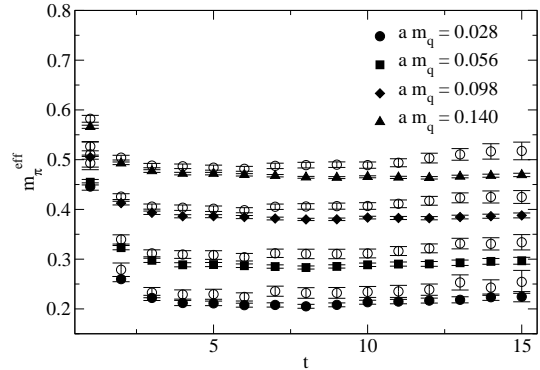}
\vspace*{-0.5cm}
  \caption{Effective pion masses from $\langle A_4(t) A_4(0)\rangle$ before 
  (solid symbols) and after (open symbols) removal of zero modes from the 
  quark propagators.}
\vspace*{-0.25cm}
  \label{fig:zeromodeproj}
\end{figure}

\begin{table}[!b]
  \begin{center}
    \begin{tabular}{|c|c|c|}
      \hline
      $a m_q$ &  $a m_\pi$ & $m_\pi$ [MeV] \\ \hline
      $0.028$ & $0.211(3)$ & $439(6)$\\
      $0.056$ & $0.288(2)$ & $599(4)$\\
      $0.098$ & $0.383(2)$ & $797(4)$\\
      $0.140$ & $0.466(2)$ & $969(4)$\\
      \hline
    \end{tabular}
  \end{center}
  \caption{Pion masses from $\langle A_4(t)A_4(0)\rangle$. We have used 
  $a=0.095$ fm to convert the lattice numbers to physical units.}
  \label{tab:mpi_a4}
\end{table}

The correlation functions $\langle P(t) P(0) \rangle$ and 
$\langle A_4(t) P(0) \rangle$ give compatible results, albeit with larger 
error bars. If one explicitly removes the zero mode contributions from the 
quark propagators one arrives at the result shown in Fig.~5. It appears that
the pion mass obtained from $\langle A_4(t) A_4(0)\rangle$ increases by 
$\approx 10\%$ ($5\%$) at the lowest (highest) quark mass. Whether this is
a valid procedure to reduce finite size effects is not clear to us, because it 
involves a nonlocal step~\cite{Hasenfratz}. But it should be taken as a
warning that finite size effects might still be large.

In Fig.~6 we plot $m_\pi^2$ against $m_q$, and in Fig.~7 we show the deviation
of $m_\pi^2$ from linearity. Quenched chiral perturbation theory predicts, in
the infinite volume, 
\begin{equation}
m_\pi^2 = A m_q \big( 1-\delta[\ln(Am_q/\Lambda_\chi^2) + 1]\big) +O(m_q^2) .
\end{equation}
We fit our data by
\begin{equation}
\label{eq:mpi_chpt}
m_\pi^2 = A m_q + Bm_q \ln m_q + C m_q^2 .
\end{equation}
The result of the fit is shown by the curves in Figs.~6 and 7. Using
$\Lambda_\chi = 4\pi f_\pi$ ($f_\pi = 93$ MeV), we derive $\delta = 0.26(9)$.
This number agrees, within error bars, with what one would expect. However, 
before one can draw any conlusions, a careful study of finite size effects 
must be performed. 

\begin{figure}[tbp]
  \centering
  \epsfig{file=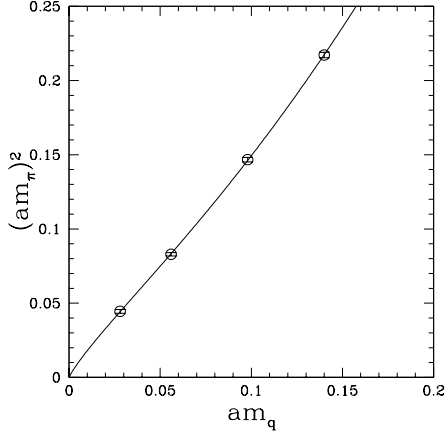, width=6cm}
\vspace*{-0.5cm}
  \caption{The pion mass $(am_\pi)^2$ as a function of $am_q$, together with 
  the fit.}
\vspace*{-0.25cm}
\end{figure}

\begin{figure}[tbp]
  \centering
  \epsfig{file=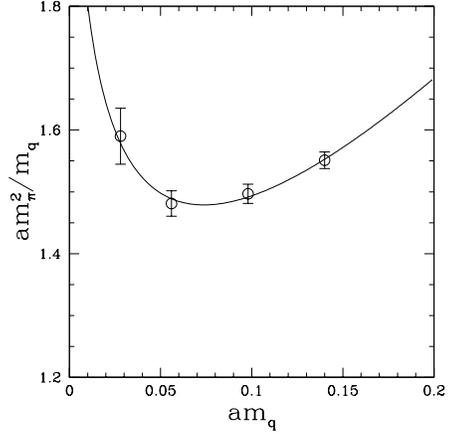, width=6cm}
\vspace*{-0.5cm}
  \caption{The ratio $am_\pi^2/m_q$ as a function of $am_q$, together with the
  fit.}
\vspace*{-0.25cm}
\end{figure}


\subsection*{\it Quark masses}
\vspace*{0.25cm}

We determine the bare light and strange quark masses, $m_\ell = (m_u + m_d)/2$
and $m_s$, from the physical pion and kaon masses~\cite{Golterman,Yoshie}:
\begin{eqnarray}
m_{\pi\,{\rm phys}}^2 &\!\!=\!\!& A\, m_\ell + B\, m_\ell \ln m_\ell 
+ C\, m_\ell^2 
, \\[0.2em]
m_{K\,{\rm phys}}^2 &\!\!=\!\!& A\, (m_\ell + m_s)/2 
+ B\, (m_\ell + m_s)/2\nonumber \\
&\!\!\times\!\!& \Big(\frac{m_s \ln m_s -m_\ell \ln m_\ell}{m_s-m_\ell}-1\Big)
 \\
&\!\!+\!\!& C\, [(m_\ell + m_s)/2]^2 . \nonumber
\end{eqnarray} 
We find $am_\ell = 0.0020(3)$ and $am_s = 0.068(2)$. The renormalized quark 
mass is given by $m_q^R = Z_m(a\mu)\, m_q$  with $Z_m = 1/Z_S$. Using 
$Z_S^{TI}(2\,{\rm GeV}) = 1.184$ (and $a=0.095$ fm), we obtain
\begin{eqnarray}
m_\ell^{\overline{MS}}(2\,{\rm GeV}) &=& 3.5(3) \,\, {\rm MeV} , \\
m_s^{\overline{MS}}(2\,{\rm GeV}) &=& 119(4)\, {\rm MeV} .
\end{eqnarray}
If we use $Z_S^{VI}(2\,{\rm GeV}) = 1.426$ instead, the numbers reduce to
\begin{eqnarray}
m_\ell^{\overline{MS}}(2\,{\rm GeV}) &=& 2.9(3) \, {\rm MeV} , \\
m_s^{\overline{MS}}(2\,{\rm GeV}) &=& 99(3)\;\, {\rm MeV} .
\end{eqnarray}
The light quark mass $m_\ell$ should not be taken too seriously, because it is
strongly affected by chiral logarithms.


\subsection*{\it Nonperturbative determination of $Z_A$ }
\vspace*{0.25cm}

The renormalization constant $Z_A$ can be computed nonperturbatively
from the Ward identity (cf.~\cite{Liu})
\begin{equation}
Z_A = \lim_{t \rightarrow \infty} \frac{2 m_q}{m_\pi} 
\frac{\langle P(t) P(0)\rangle}{\langle A_4(t) P(0)\rangle} .   
\label{zanonpert}
\end{equation}
The result is plotted in Fig.~8. We see that $Z_A$ depends only weakly on
the quark mass. A linear extrapolation to the chiral limit gives 
\begin{equation}
\label{za}
Z_A=1.416(2) .
\end{equation}
This number lies  $20\%$ above the tadpole improved 
perturbative result.

\begin{figure}[tbp]
  \centering
  \epsfig{file=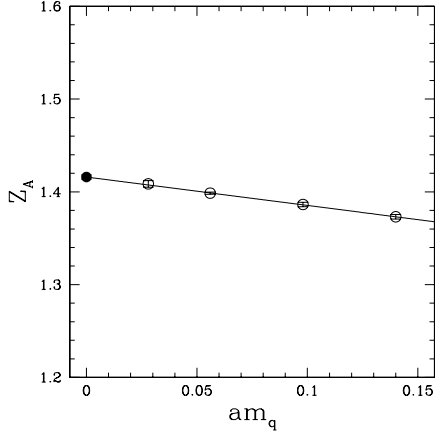,width=6cm}
\vspace*{-0.5cm}
  \caption{The renormalization constant $Z_A$ as a function of $am_q$, 
  together with the chiral extrapolation.}
\vspace*{-0.25cm}
  \label{fig:za}
\end{figure}  

\begin{table}[!b]
  \begin{center}
    \begin{tabular}{|c|l|c|}
      \hline
      $a m_q$ &  \multicolumn{1}{c|}{$a m_N$} & $m_N$ [MeV]\\ \hline
      $0.028$ & $0.60(3)$  & $1250(60)$\\
      $0.056$ & $0.65(1)$  & $1350(20)$\\
      $0.098$ & $0.775(10)$  & $1612(20)$\\
      $0.140$ & $0.875(5)$ & $1820(10)$\\
      \hline
    \end{tabular}
  \end{center}
  \caption{Nucleon masses. We have used $a=0.095$ fm to convert the lattice
  numbers to physical units.}
  \label{tab:mn}
\end{table}

\subsection*{\it Nucleon mass}
\vspace*{0.25cm}

Our results for the nucleon mass are given in Table~3. In Fig.~9 we plot 
$(m_N r_0)^2$ against $m_q$. Except for the lowest mass, which currently
is not very well determined, the data points lie on a straight line, which 
extrapolates surprisingly well to the experimental value.  

\begin{figure}[tbp]
  \centering
  \epsfig{file=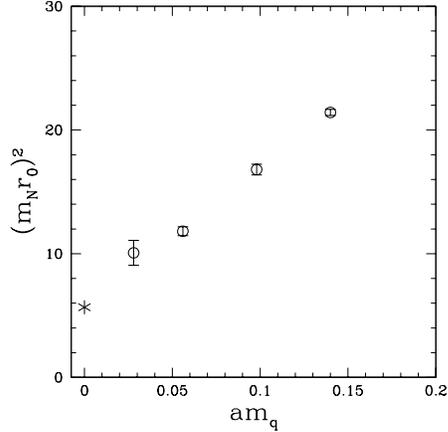, width=6cm}
\vspace*{-0.5cm}
  \caption{The nucleon mass $(m_N r_0)^2$ as function of $am_q$, together with 
 the experimental result ($\ast$), using $r_0/a = 5.26$.}
\vspace*{-0.25cm}
  \label{fig:mn_vs_mq }
\end{figure}



\section{NUCLEON MATRIX ELEMENTS}

While hadron masses are determined from two-point correlation
functions, nucleon matrix elements of quark bilinear operators $\mathcal{O}$,
\begin{equation}
\langle N|\mathcal{O}|N\rangle , \; \langle N| N\rangle = 2 m_N,
\end{equation}
are derived from ratios of three-point to two-point functions,
\begin{equation}
   R \equiv \frac{\langle N(t) \mathcal{O}(\tau) \bar{N}(0)\rangle}
            {\langle N(t) \bar{N}(0)\rangle}
     \simeq    \frac{1}{2m_N} \langle N | \mathcal{O} | N \rangle ,
\label{ratio}
\end{equation}
which for $t \gg \tau \gg 0$ are proportional to the desired
matrix element, as shown on the r.h.s. of eq.~(\ref{ratio}). Here $N(t)$ is a 
suitable baryon operator. We consider nucleons of zero momentum only.  
This technique is by now standard. For details and the numerical 
implementation the reader is referred to~\cite{QCDSF4,Best}. 
The calculation of three-point functions requires additional inversions and  
therefore is computationally expensive. The calculations in this Section are 
based on $O(50)$ configurations, so the results must be regarded as very
preliminary.

While hadron masses are automatically $O(a)$ improved, operator matrix
elements are generally not, except in the chiral limit. However, in 
distinction to improved Wilson fermions, improvement for overlap fermions is 
very simple and universal~\cite{QCDSF1}:
\begin{equation}
\mathcal{O}^{\rm imp} = \Big(1-\frac{am_q}{2\rho}\Big)^{-1} \, \mathcal{O}.
\end{equation}
This a major advantage, as we do not have to compute matrix 
elements of higher dimensional operators, nor their associated improvement
coefficients. 


\begin{figure}[!t]
\begin{center}
\epsfig{figure=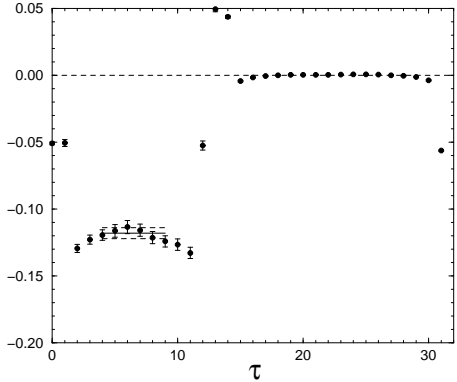,width=6cm}
\vspace{-0.5cm}
\caption{The ratio $R$ plotted against the position $\tau$ of the operator
$\mathcal{O}_b$ for $am_q = 0.14$. The source ($\bar{N}$) is placed at time 
slice $t=0$, and the sink ($N$) at $t=13$. From a plateau at $t \gg \tau \gg 
0$ we can determine $R$.} 
\end{center}
\vspace*{-0.75cm}
\end{figure}

We shall consider two operators, the local vector current $V_\mu$ and the
one-link operator $\mathcal{O}_b$. The nucleon matrix element of 
$\mathcal{O}_b$ yields the first moment of the unpolarized nucleon structure 
function. The results presented in this Section refer to improved, nonsinglet 
operators.

A typical ratio $R$ is shown in Fig.~10 for the operator $\mathcal{O}_b$
from which we can find the bare nucleon matrix element.

\subsection*{\it Local vector current}
\vspace*{0.25cm}

Let us first look at the local vector current. Due to charge conservation
\begin{equation}
\langle N|V_\mu^R |N\rangle = Z_V\,\langle N|V_\mu |N\rangle = 2 m_N ,
\end{equation}
which may be used to compute $Z_V$~\cite{QCDSF5}. We expect $Z_V = Z_A$, and 
the idea is to test this hypothesis. In Fig.~11 we plot $Z^V$
against $am_q$. We compare this result with $Z_A$ computed from 
eq.~(\ref{zanonpert}) in Section 5. We see that $Z_V$ approaches $Z_A$ 
in the chiral limit, while at our largest quark mass the numbers differ by
a few percent. Though the operators are $O(a)$ improved, discretization errors
$\sim (a m_\pi)^2$ are still possible, which might explain the small slope.
A linear extrapolation to the chiral limit gives 
\begin{equation}
Z_V=1.426(7). 
\end{equation}

\subsection*{\it First moment of the structure function}
\vspace*{0.25cm}

Let us now turn to the operator $\mathcal{O}_b$. The nucleon matrix element of
the operator $\mathcal{O}_a$ is much harder to compute, as it requires a
nonzero nucleon momentum. The matrix element of the renormalized operator
$\mathcal{O}_b$ gives~\cite{QCDSF4}
\begin{eqnarray}
\langle N|\mathcal{O}_b^R |N\rangle &=& Z_b\,
\langle N|\mathcal{O}_b |N\rangle \nonumber \\[-0.3em]
& & \\[-0.3em]
&=& - 2 m_N^2 \, \langle x \rangle \nonumber.
\end{eqnarray}
For the tadpole improved renormalization constant in the $\overline{MS}$ 
scheme we find $Z_b^{\overline{MS}}(2\,{\rm GeV}) = 1.314$. In Fig.~12 we
show the first moment of the nonsinglet nucleon structure function 
$\langle x\rangle$ in the $\overline{MS}$ scheme at $\mu = 2\,{\rm GeV}$. 
The numbers are in surprisingly good agreement with the experimental value. 
No nonanalytic behavior is needed to reconcile experimental and theoretical 
results. Previous calculations using Wilson fermions gave much larger 
values.

\begin{figure}[t]
\begin{center}
\epsfig{figure=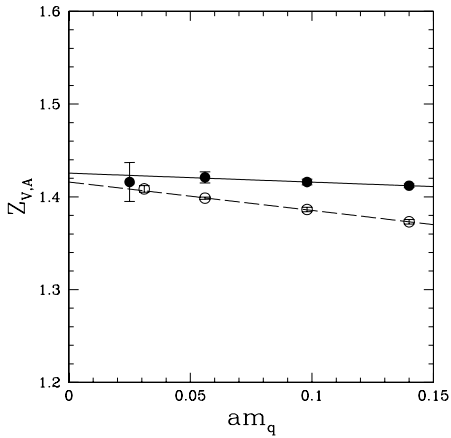,width=6cm}
\vspace{-0.5cm}
\caption{The renormalization constant $Z_V$ from this calculation (solid 
symbols) compared with $Z_A$ from Section 5 (open symbols) as a function of
$am_q$, together with linear extrapolations to the chiral limit. The symbols 
at the lowest quark mass are displaced by a small amount so that they do not 
overlap.} 
\end{center}
\vspace*{-0.75cm}
\end{figure}

\begin{figure}[t]
\begin{center}
\epsfig{file=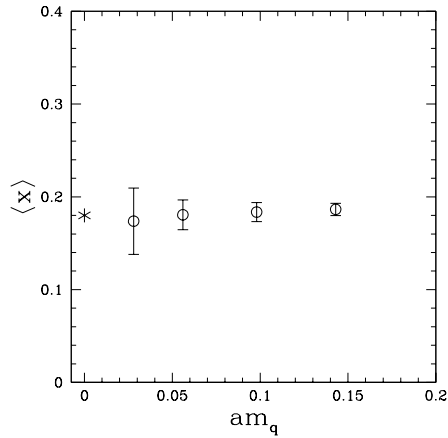,width=6cm,clip=}
\vspace*{-0.5cm}
\caption{The first moment of the nucleon structure function $\langle x\rangle$
at $\mu = 2$ GeV as a function of $am_q$. Also shown is the phenomenological
value ($\ast$).}
\end{center}
\vspace*{-0.5cm}
\end{figure}

\section{SUMMARY}

Overlap fermions have many advantages over Wilson and staggered fermions.
They provide an implementation of lattice fermions with exact 
chiral symmetry, even at finite lattice spacing. In addition, they are 
automatically $O(a)$ improved, and the task of operator renormalization is 
greatly reduced.

Calculations with overlap fermions on fine grained, phenomenologically 
relevant lattices are progressing rapidly. They are computationally costly, 
but by using an improved gauge field action and projecting out the lowest
lying eigenvalues the condition number can be substantially reduced.
We have tested the predictions of random matrix theory and find good agreement
with the unfolded distributions of the smallest eigenvalue in topological
sectors up to $|Q| = 5$. The renormalization constants of local and one-link
operators have been computed perturbatively for the tadpole improved
L\"uscher-Weisz gauge field action. The one-loop corrections turn out to be 
relatively large, and in most cases significantly larger than in the improved
Wilson fermion case~\cite{QCDSF2}, which calls for a nonperturbative 
determination. 
As a first step, we have computed $Z_A$ and $Z_V$ nonperturbatively. We find 
that $Z_A$ and $Z_V$ agree in the chiral limit, as expected. The 
nonperturbative numbers turn out to lie $20\%$ higher than the tadpole 
improved perturbative values (at $a=0.095$ fm). On the phenomenological side
we have computed the topological susceptibility and the chiral condensate from
the spectrum of low-lying eigenvalues. The topological susceptibility is 
found to be in good agreement with the Witten-Veneziano formula
\begin{equation}
m_{\eta^\prime}^2 = \frac{2 N_f}{f_\pi^2} \chi_{\rm top} ,
\label{WV}
\end{equation}
giving $m_{\eta^\prime} = 920(50)$ MeV. We employed random matrix theory 
to derive the chiral condensate in the infinite volume. We went on to compute 
the pion,
nucleon and quark masses. We find some signal for chiral logarithms. Both
$r_0/a$~\cite{Gattringer}, using $r_0=0.5$ fm, and the nucleon mass give 
compatible values for the lattice spacing. Our lowest pion mass so far is
$m_\pi \approx 400$ MeV. There is quite a big uncertainty in the strange 
quark mass due to an uncertainty in $Z_m$. Finally, we were able to compute
the first moment of the nucleon structure function. Good agreement with the
experimental value has been found.

\section*{ACKNOWLEDGEMENTS}

The numerical calculations have been performed at NIC J\"ulich, ZIB Berlin 
and NeSC Edinburgh (IBM Regatta), NIC Zeuthen and Southampton (PC Cluster), 
and HPCF Cranfield (SunFire). We thank these institutions
for support. This work is supported by the European Community's Human Potential
Program under contract HPRN-CT-2000-00145 Hadrons/Lattice QCD and by DFG 
under contract FOR 465 (Forschergruppe Gitter-Hadronen-Ph\"anomenologie).


\end{document}